\documentclass[conference, letterpaper]{./IEEEtran}

\ifCLASSINFOpdf
  \usepackage[pdftex]{graphicx}
\else
 
\fi

\usepackage{amsmath}
\usepackage{amssymb}
\usepackage{amsfonts}
\usepackage{subfig}
\usepackage{multirow}
\usepackage{multicol}
\usepackage{subfig}
\usepackage{flushend}

\begin{document}

\title{Optimal DVB-S2 Spectral Efficiency with Hierarchical Modulation}

\author{
\IEEEauthorblockN{Hugo M{\'e}ric}
\IEEEauthorblockA{NIC Chile Research Labs\\
Santiago, Chile\\
Email: hmeric@niclabs.cl}
}

\maketitle

\begin{abstract}
We study the design of a DVB-S2 system in order to maximise spectral efficiency. This task is usually challenging due to channel variability. The solution adopted in modern satellite communications systems such as DVB-SH and DVB-S2 relies mainly on a time sharing strategy. Recently, we proposed to combine time sharing with hierarchical modulation to increase the transmission rate of broadcast systems. However, the optimal spectral efficiency remained an open question. In this paper, we show that the optimal transmission rate is the solution of a linear programming problem. We also study the performance of the optimal scheme for a DVB-S2 use case. 
\end{abstract}

\IEEEpeerreviewmaketitle

\section{Introduction}

In most broadcast applications, the signal-to-noise ratio (SNR) experienced by each receiver can be quite different. For instance, in satellite communications the channel quality decreases with the presence of clouds in Ku or Ka band, or with shadowing effects of the environment in lower bands.

\begin{figure}[!t]
\centering
\includegraphics[width = 0.75\columnwidth]{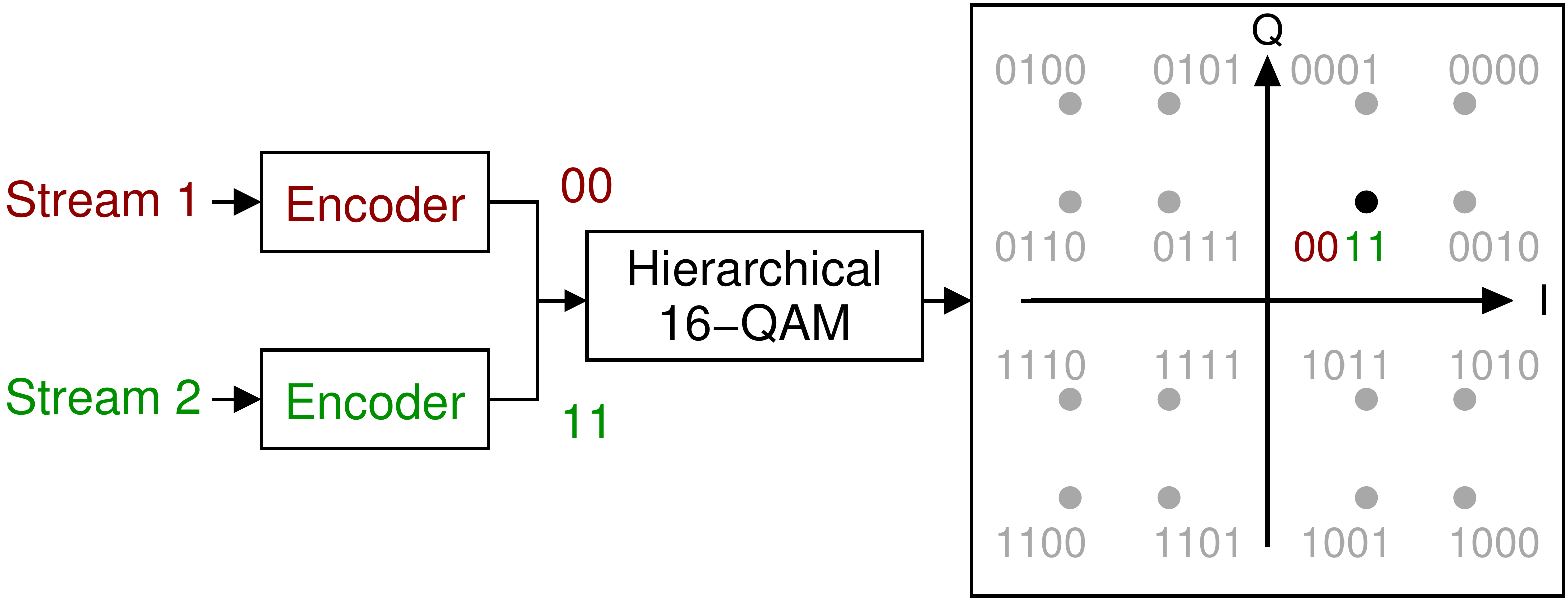}
\caption{Two-layers hierarchical modulation using a 16-QAM}
\label{hm_principle}
\vspace{-3mm}
\end{figure}

The first solution for broadcasting was to design the system for the worst-case reception, but this leads to poor performance as many receivers do not exploit their full potential. Two other schemes were then proposed: time division multiplexing with variable coding and modulation, and superposition coding \cite{cover, bergmans}. Time division multiplexing, or time sharing, allocates a proportion of time to communicating with each receiver using any modulation and error protection level. This functionality, called variable coding and modulation (VCM), is in practice the most used in standards today. If a return channel is available, VCM may be combined with adaptive coding and modulation (ACM) to optimise the transmission parameters (code rate and modulation) \cite{s2}. In superposition coding, the available energy is shared among several service flows which are sent simultaneously in the same band. This scheme was introduced by Cover in order to improve the transmission rate from a single source to several receivers \cite{cover}. When communicating with two receivers, the principle is to superimpose information for the receiver with the best SNR. This superposition can be done directly at the forward error correction level or at the modulation level as shown in \figurename~\ref{hm_principle} with a 16 quadrature amplitude modulation (16-QAM).

Hierarchical modulation is a practical implementation of superposition coding. Although hierarchical modulation has been introduced to improve throughput of broadcast channels, it has many other applications such as broadcasting local content \cite{local_content}, providing unequal protection \cite{svc_hm}, improving the performance of relay communication system \cite{relaycom} or backward compatibility \cite{broad05}.

Our work focuses on using hierarchical modulation in modern broadcast systems to increase the transmission rate. For instance, even if the DVB-S2\footnote{Digital Video Broadcasting - Satellite - Second Generation} low-density parity-check codes approach the Shannon limit for the additive white Gaussian noise (AWGN) channel with one receiver \cite{SAT:SAT787}, the throughput can be greatly increased for the broadcast case. Indeed, we recently showed that combining ACM with hierarchical modulation improves the spectral efficiency of a DVB-S2 system \cite{broad13,icc14}. To that end, we introduced hierarchical versions of the modulations considered in the DVB-S2 standard, i.e., the quadrature phase-shift keying (QPSK), the 16 and 32 amplitude and phase-shift keying modulations (16-APSK and 32-APSK). The performance improvement is significant (up to 10\% in some cases) for a large range of channel quality when the receivers experience low or large SNR. 

When combining time sharing with hierarchical modulation, the optimal spectral efficiency remained an open question. In this paper, we show that the optimal transmission rate is the solution of a linear program. This is the main contribution of our work. Then we compare the performance of the optimal scheme with a reference system without hierarchical modulation and with the suboptimal scheme proposed in \cite{icc14}.
 
The paper is organised as follows: Section~\ref{part2} gives a short introduction to hierarchical modulation. Section~\ref{part3} presents how to obtain the optimal spectral efficiency by solving a linear programming problem. We evaluate the performance of the optimal scheme in Section~\ref{part4}. Section~\ref{part5} is dedicated to related work and discussion. Finally,  we conclude the paper by summarising the results and presenting the future work in Section~\ref{part6}.

\textbf{Notations.} Column vectors appear in bold and $\cdot^t$ denotes the transpose operation throughout this paper.

\section{Short introduction to hierarchical modulation}\label{part2}

\begin{figure*}[!t]
\centering
\includegraphics[width = 1.925\columnwidth]{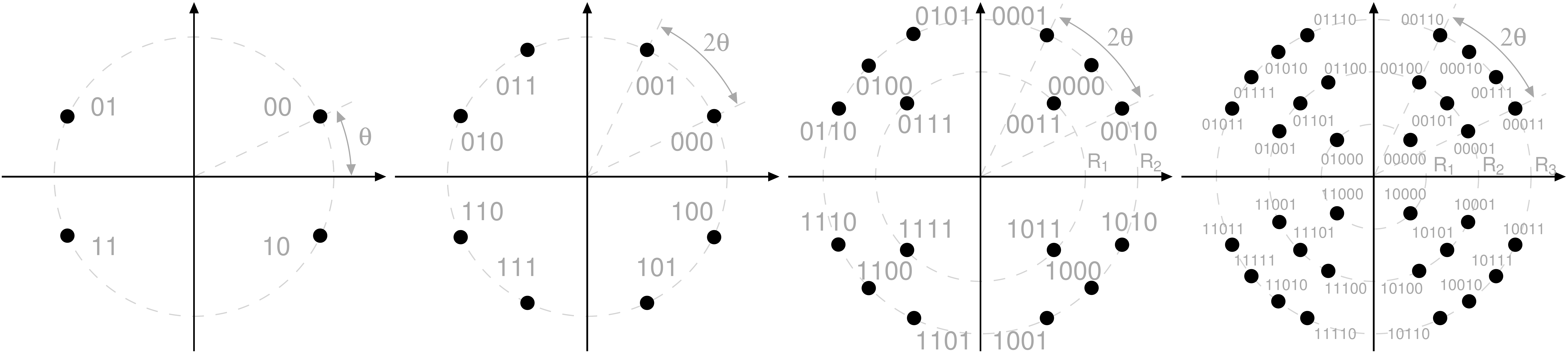}
\caption{Non-uniform constellations used in this work. From left to right: QPSK, 8-PSK, 16-APSK and 32-APSK.}
\label{nu_constellations}
\end{figure*}

As already mentioned, hierarchical modulations merge several streams in a same symbol. The principle is to share the available energy between each stream. In this paper, we consider 2-layers hierarchical modulations where two streams are merged in each constellation symbol (see \figurename~\ref{hm_principle}). The extension to $n$-layers hierarchical modulation is straightforward.

As each stream usually does not use the same amount of energy, hierarchical modulations often rely on non-uniform constellations where the symbols are not uniformly distributed in the space as depicted in \figurename~\ref{hm_principle}. Our use case in Section~\ref{part4} relies on the DVB-S2 standard that considers the following constellations: QPSK, 8-PSK, 16-APSK and 32-APSK. Except for the hierarchical 8-PSK that was already considered in the standard for backward compatibility purpose, we recently introduced a hierarchical version for each of the previous modulations \cite{broad13, icc14}. \figurename~\ref{nu_constellations} presents the respective constellations and mappings.

The geometry of non-uniform constellations is described using the constellation parameter(s). The non-uniform QPSK and 8-PSK require one parameter $\theta$. For the 16-APSK, the constellation parameters are the ratio between the radius of the outer ($R_2$) and inner ($R_1$) rings $\gamma = R_2 / R_1$, and the half angle between the points on the outer ring in each quadrant $\theta$. Finally, the 32-APSK is described using three parameters: the ratio between the radius of the middle ($R_2$) and inner ($R_1$) rings $\gamma_1 = R_2 /R_1$, the ratio between the radius of the outer ($R_3$) and inner ($R_1$) rings $\gamma_2 = R_3 / R_1$ and the half angle between the points on the outer ring in each quadrant $\theta$. 

We determine the previous constellation parameters by exploiting their relation with the amount of energy allocated to the transmitted streams \cite{broad13, icc14}. We provide all the parameters used in the simulations in Section~\ref{part4}. A detailed description for designing the hierarchical 16-APSK can be found in \cite{broad13}, while the hierarchical QPSK and 32-APSK are treated in \cite{icc14}.

As already stated, each transmitted symbol merges data from two streams (noted stream 1 and stream 2). Thus some bits in the symbol mapping are assigned to carry stream 1, while the others carry stream 2. Table~\ref{assignment} resumes the bits assigned to each stream for the hierarchical modulations depicted in \figurename~\ref{nu_constellations}. It is important to mention that each receiver decodes indifferently stream 1 or 2 in our framework.
\begin{table}[ht!]
\renewcommand{\arraystretch}{1.1}
\caption{Bits assignment to transmit stream 1 and 2}
\label{assignment}
\centering
\begin{tabular}{c|c|c|c} 
\hline
\textbf{Hierarchical} & \textbf{Symbol}		& \textbf{Bit(s) assigned}     & \textbf{Bit(s) assigned} \\
\textbf{modulation}   & \textbf{mapping}	& \textbf{to stream 1}	       &  \textbf{to stream 2}	  \\\hline
QPSK 		      & $b_1b_2$          	& $b_1$             	       & $b_2$ 			  \\\hline
8-PSK 		      & $b_1b_2b_3$       	& $b_1b_2$          	       & $b_3$ 			  \\\hline
16-APSK 	      & $b_1b_2b_3b_4$    	& $b_1b_2$        	       & $b_3b_4$ 		  \\\hline
32-APSK 	      & $b_1b_2b_3b_4b_5$ 	& $b_1b_2$        	       & $b_3b_4b_5$ 		  \\\hline 
\end{tabular}
\end{table}

\section{Optimal spectral efficiency}\label{part3}

\textbf{Definitions and hypotheses.} We study a broadcast channel with $n$ receivers. We assume that the system implements hierarchical and non-hierarchical modulations. Moreover, we only consider 2-layers hierarchical modulation in this paper. The extension to $n$-layers hierarchical modulation will be discussed in Section~\ref{part5}. 

We define a rate vector $\mathbf{r} = (r_1, \dots, r_n)^t \in \mathbb{R}^n$ where $r_i$ represents the spectral efficiency of receiver $i$ ($1\leqslant i \leqslant n$). A rate vector $\mathbf{r}$ is said to be achievable if there exists some transmission parameters such that receiver $i$ has a spectral efficiency equals to $r_i$ ($1\leqslant i \leqslant n$).

Non-hierarchical modulations enable rate vectors of the form $(0 \dots 0, r_k, 0 \dots 0)^t$ ($1\leqslant k \leqslant n$) with only one non-zero component, while 2-layers hierarchical modulations provide rate vectors of the form $(0 \dots 0, r_i, 0 \dots 0, r_j, 0 \dots 0)^t$ ($1\leqslant i,j \leqslant n$) with two non-zero components. The transmission rates $r_i$, $r_j$ and $r_k$ depend on the SNR of the receivers, but also the modulations and code rates available in the system.

Considering a set of $k$ achievable rate vectors $(\mathbf{r_i})_{1\leqslant i \leqslant k}$, time sharing enables to achieve any transmission rate
\begin{equation}
\mathbf{r} = \sum_{i=1}^{k} t_i \mathbf{r_i},
\end{equation} 
where $t_i \geqslant 0$ ($1\leqslant i \leqslant k$) and $\sum_i t_i = 1$. We refer to the set $(t_i)_{1\leqslant i \leqslant k}$ as the time sharing coefficients.

Finally, we are interested in offering the same time-averaged spectral efficiency to all the receivers. In other words, we seek to obtain a rate vector of the form
\begin{equation}
\mathbf{r} = \sum_{i=1}^{k} t_i \mathbf{r_i} = (R, \dots, R),
\end{equation}
where $R \geqslant0$ is the time-averaged spectral efficiency. We will explain how to include rate constraints between the receivers in Section~\ref{part5}.

\textbf{Problem formulation.} Given the (hierarchical and non-hierarchical) modulations and the code rates defined in the system, there is a finite (possibly large) number of achievable rate vectors $(\mathbf{r_i})_{1\leqslant i \leqslant k}$ with one or two non-zero coefficients. The problem is to find a set of time sharing coefficients $(t_i)_{1\leqslant i \leqslant k}$ in order to maximise the time-averaged spectral efficiency $R$. More formally, we solve
\begin{equation}
\begin{aligned}
& \underset{(t_i)_{1\leqslant i \leqslant k}}{\text{maximise}}   & & R \\
& \text{subject to} 						   & & \sum_{i=1}^{k} t_i \mathbf{r_i} = (R, \dots , R)^t, \\
&		    						   & & \sum_{i=1}^{k} t_i = 1, \\
&		    						   & & t_i \geqslant 0 \text{ } (1 \leqslant i \leqslant k).
\end{aligned}
\label{optim1}
\end{equation}

Eq.~(\ref{optim1}) is a linear programming problem that can be solved with classical solvers such as MOSEK, a high performance software for large-scale optimisation problems \cite{mosek}. We now present how to express (\ref{optim1}) in a canonical form that serves as an input for the usual linear programming solvers.

\textbf{Linear program.} To begin with, we define the vectors
\begin{equation}
\mathbf{x} = \left( t_1, \dots , t_k, R\right)^t \in \mathbb{R}^{k+1}
\end{equation}
and
\begin{equation}
\mathbf{c} = \left( 0 , \dots , 0 , 1 \right)^t \in \mathbb{R}^{k+1}.
\end{equation}
The vector $\mathbf{x}$ contains the time sharing coefficients and the time-averaged spectral efficiency. Therefore we have the following relationship
\begin{equation}
R = \mathbf{c}^t\mathbf{x}.
\end{equation}

Now, we seek to express the constraints in (\ref{optim1}) in a matrix form. To that end, we introduce the matrix
\begin{equation}
A =
\left(\begin{array}{ccc|c}
      	                &       &     	                & -1     \\
  \mathbf{r}_\mathbf{1} & \dots & \mathbf{r}_\mathbf{k} & \vdots \\
                        &       &    	                & -1     \\\hline
  1                     & \dots & 1   	                & 0

\end{array}\right) \in \mathbb{R}^{(n+1)\times(k+1)}
\label{A}
\end{equation}
and the vector
\begin{equation}
\mathbf{b} = \left( 0 , \dots , 0 , 1 \right)^t \in \mathbb{R}^{n+1}.
\end{equation}
As a result, the first two constraints in (\ref{optim1}) are equivalent to $A\mathbf{x} = \mathbf{b}$. The last constraint is simply expressed as $\mathbf{x} \geqslant 0$ (component-wise inequality).

Finally, the previous definitions enable to rewrite (\ref{optim1}) as
\begin{equation}
\begin{aligned}
& \underset{\mathbf{x}}{\text{maximise}} & & \mathbf{c}^t\mathbf{x} \\
& \text{subject to}			 & & A\mathbf{x} = \mathbf{b}, \\
&					 & & \mathbf{x} \geqslant 0,
\end{aligned}
\label{canonical_form}
\end{equation}
which is a linear programming problem in standard form that existing solvers can solve efficiently. The solver output is the vector $\mathbf{x_{opt}}$ that contains all the necessary information for the service provider, especially the time sharing coefficients resulting in the optimal spectral efficiency.

\section{Performance evaluation}\label{part4}

We present in this part the simulations setup (scenarios, channel model, etc) and the results.

\textbf{Scenarios.} Our simulations involve the code rates in the DVB-S2 standard and the following modulations: QPSK, 8-PSK, 16-APSK, 32-APSK and their hierarchical versions.

We compare the performance of three different schemes based on DVB-S2. The reference scenario is equivalent to the standard and only transmits with the non-hierarchical modulations. The two other scenarios employ hierarchical modulations. We consider the suboptimal solution proposed in \cite{icc14}. More details concerning this scenario are given in Section~\ref{part5}. The last scheme relies on linear programming to obtain the optimal spectral efficiency.

Table~\ref{constellation_parameters} resumes the constellation parameters for the non-uniform constellations used in our simulations. Compared to the standard, we add 22 hierarchical modulations. The error performance of the non-hierarchical modulations are summarised in \cite[Table 13]{dvbs2}, while the hierarchical modulations performance can be found in \cite{broad13} and \cite{icc14}.
\begin{table}[ht!]
\renewcommand{\arraystretch}{1.1}
\caption{Constellation parameters ($\theta$ is in degree; $\gamma$, $\gamma_1$ and $\gamma_2$ are dimensionless)}
\label{constellation_parameters}
\centering
\begin{tabular}{c|l} 
\hline
\textbf{QPSK}                  & \multirow{2}{*}{$45^\circ; 42^\circ; 39^\circ; 36^\circ; 33^\circ; 30^\circ; 27^\circ; 24^\circ; 18^\circ$} \\
$\theta$         	       & 								        				     \\\hline
\textbf{8-PSK}                 & \multirow{2}{*}{$30^\circ; 27^\circ; 24^\circ; 18^\circ$} 	      					     \\
$\theta$         	       & 								        				     \\\hline 
\textbf{16-APSK}               & \multirow{2}{*}{$(31.5^\circ, 2.8); (28.4^\circ, 2.3); (25.1^\circ, 1.9); (20.9^\circ, 1.6)$}		     \\
$(\theta, \gamma)$ 	       & 								      					     \\\hline 
\textbf{32-APSK}  	       & $(32.3^\circ, 2.4, 5); (30.2^\circ, 1.8, 3.4); (28.4^\circ, 1.6, 2.6);$  	      			     \\
$(\theta, \gamma_1, \gamma_2)$ & $(25.6^\circ, 1.6, 2.2); (17.4^\circ, 1.8, 2.4)$  		    	      				     \\\hline 
\end{tabular}
\end{table}

\textbf{Link unavailability.}
In our work, we focus on optimising the spectral efficiency. However the link unavailability, defined as the percentage of receivers that can not decode any stream, is another important metric for broadcast systems. Indeed, the transmission parameters maximising the spectral efficiency may also produce a small coverage. A trade-off (defined by the operator) exists between a high spectral efficiency and a reasonable unavailability.

Considering the reference scheme, the receivers with a SNR under a given threshold (-2.35 dB as shown in \cite[Table 13]{dvbs2}) are not able to decode any stream. These receivers are not taken into account in the two other schemes to ensure the same coverage for all the scenarios. This enables a fair comparison where the coverage remains identical and only the transmission rate varies depending on the scenarios.

\begin{figure}[!ht]
\centering
\includegraphics[width = 0.8\columnwidth]{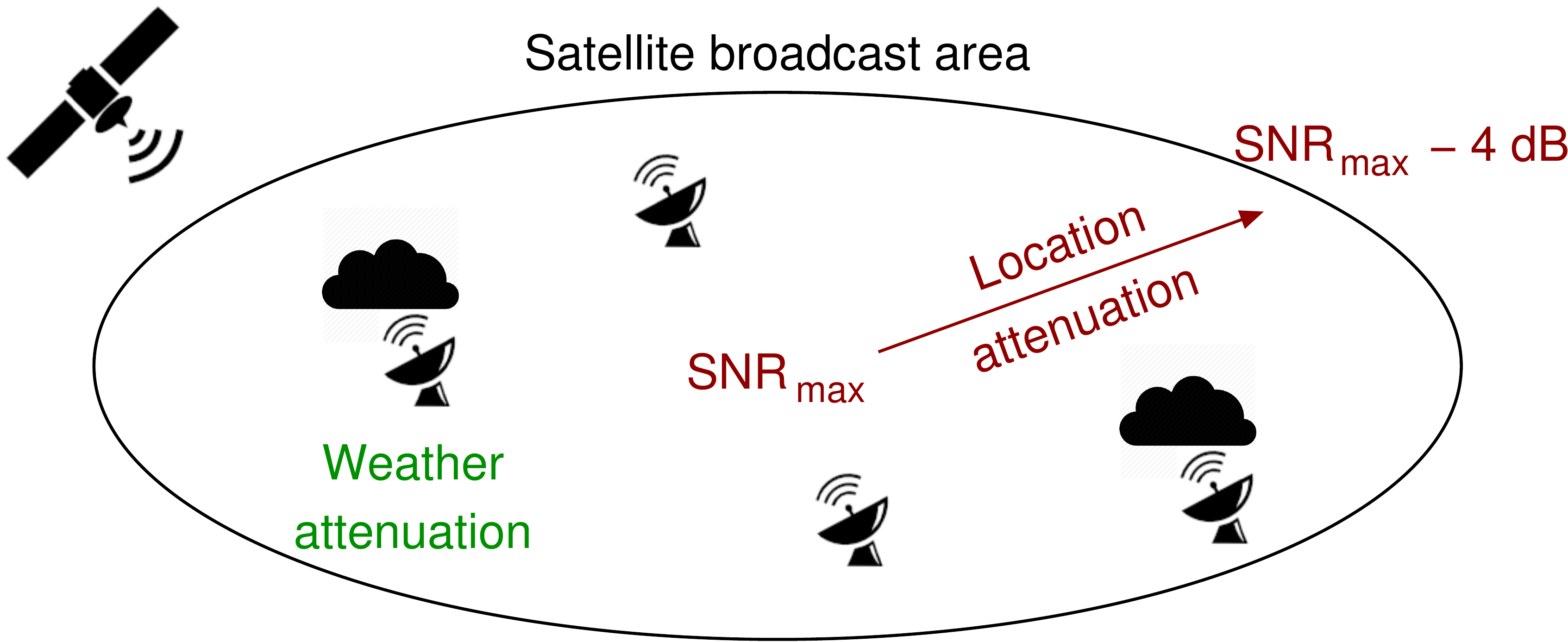}
\caption{Attenuation sources in a broadcast area: location of the terminal with respect to the center of coverage and weather.}
\label{spot}
\end{figure}

\textbf{Channel model.} We study the set of receivers located in a given spot beam of a geostationary satellite broadcasting in the Ka band. The transmission is subject to AWGN. Our channel model takes into account two sources of attenuation as depicted in \figurename~\ref{spot}: the relative location of the terminal with respect to the center of (beam) coverage and the weather. 

Concerning the attenuation due to the location, the idea is to set $\text{SNR}_{\text{max}}$, the SNR at the center of the spot beam (see \figurename~\ref{spot}), and use the radiation pattern of a parabolic antenna to model the attenuation. An approximation of the radiation pattern is 
\begin{equation}
G(\theta) = G_{\text{max}} \left( 2\frac{J_1 \left( \sin(\theta) \frac{\pi D}{\lambda} \right)}{\sin(\theta) \frac{\pi D}{\lambda}} \right)^2,
\label{eq_rayonnement}
\end{equation}
where $J_1$ is the first order Bessel function, $D$ is the antenna diameter and $\lambda=c/f$ is the wavelength \cite{antenna}. Our simulations use $D=1.5\text{ m}$ and $f=20 \text{ GHz}$. Moreover, we consider a typical multispot system where the edge of each spot beam is 4 dB below the center of coverage as shown in \figurename~\ref{spot}. Finally, we obtain the (location) attenuation distribution assuming a uniform repartition of the population inside the broadcast area.

\begin{figure}[!ht]
\centering
\includegraphics[width = 0.9\columnwidth]{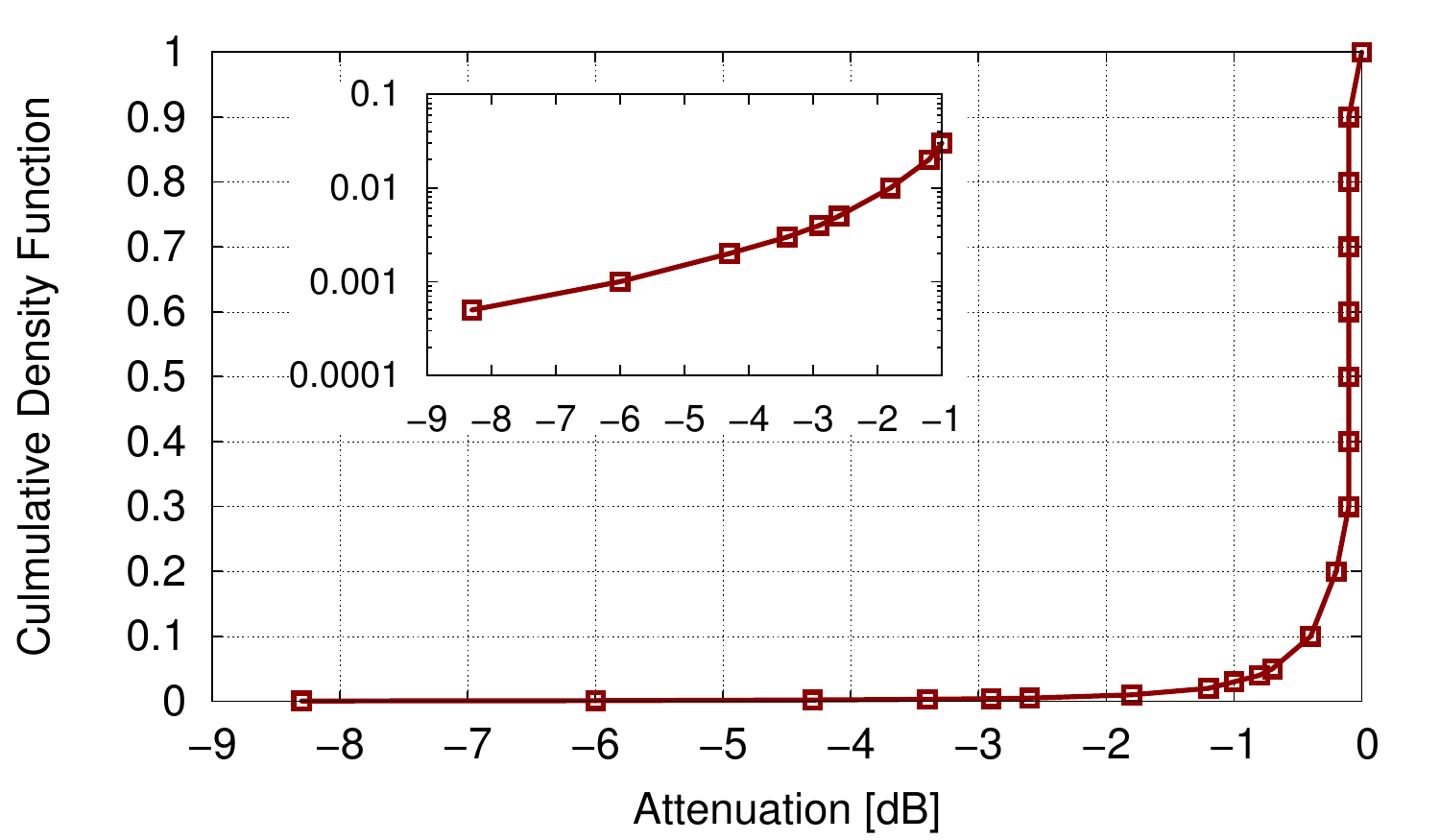}
\caption{Attenuation distribution due to weather for the broadcasting satellite service band (provided by the CNES)}
\label{distrib_s2}
\end{figure}

The weather attenuation is drawn according to the attenuation distribution of the broadcasting satellite service band depicted in \figurename~\ref{distrib_s2}, provided by the centre national d'\'etudes spatiales (CNES). More precisely, it is a temporal distribution for a given location in Toulouse, France. In this paper, we assume that the SNR distribution for the receivers in the beam coverage at a given time is equivalent to the temporal distribution at a given location.

Finally, our model combines the two attenuations previously described to estimate the SNR distribution. From a set of receivers, we first compute the attenuation due to the location. Then, for each receiver we draw the attenuation caused by the weather according to the distribution provided by the CNES.

\textbf{Additional information.} 
We remind that we seek to offer the same time-averaged spectral efficiency to all the receivers. 

Each simulation requires two input parameters: the number of receivers and $\text{SNR}_{\text{max}}$ (the SNR at the center of the coverage area). We consider a broadcast area of 500 receivers and we vary $\text{SNR}_{\text{max}}$ between 2 and 21 dB. Considering the reference scheme and almost no weather attenuation, we have the following facts: when $\text{SNR}_{\text{max}}$ is equal to 2 dB, the receivers on the edge of the spot beam (that suffer a location  attenuation of 4 dB) experience a SNR of -2 dB and are still able to decode the QPSK modulated signal with the lowest code rate. On the contrary, when $\text{SNR}_{\text{max}}$ is equal to 21 dB, the receivers on the edge are able to decode the 32-APSK with the highest code rate resulting in the best transmission rate.

The SNR value of each receiver is drawn according to the distribution presented above. This SNR is fixed over all times for a given simulation. We also assume that the transmitter has knowledge of the SNR at the receivers. In practice, this corresponds to a system that implements ACM.

Concerning the implementation, we use Matlab to run the simulations. We interface Matlab with MOSEK to speed up the simulations. The function \texttt{proglin} solves the linear program expressed in (\ref{canonical_form}).

\begin{figure*}[!ht]
\centerline{\subfloat[Hierarchical QPSK (9 constellations)]{\includegraphics[width=0.45\textwidth]{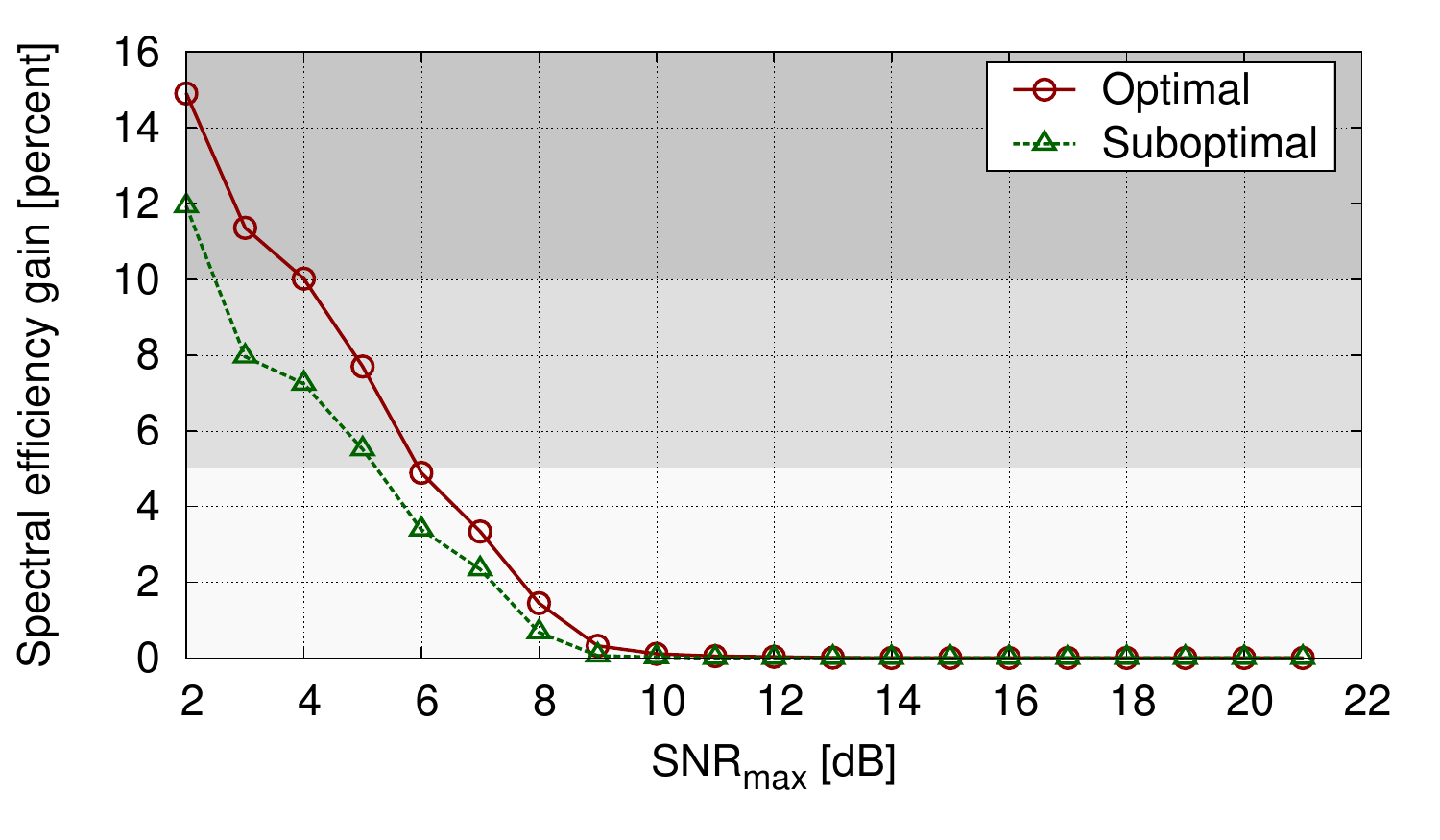}%
\label{qpsk}}%
\hfil
\subfloat[Hierarchical 8-PSK (4 constellations)]{\includegraphics[width=0.45\textwidth]{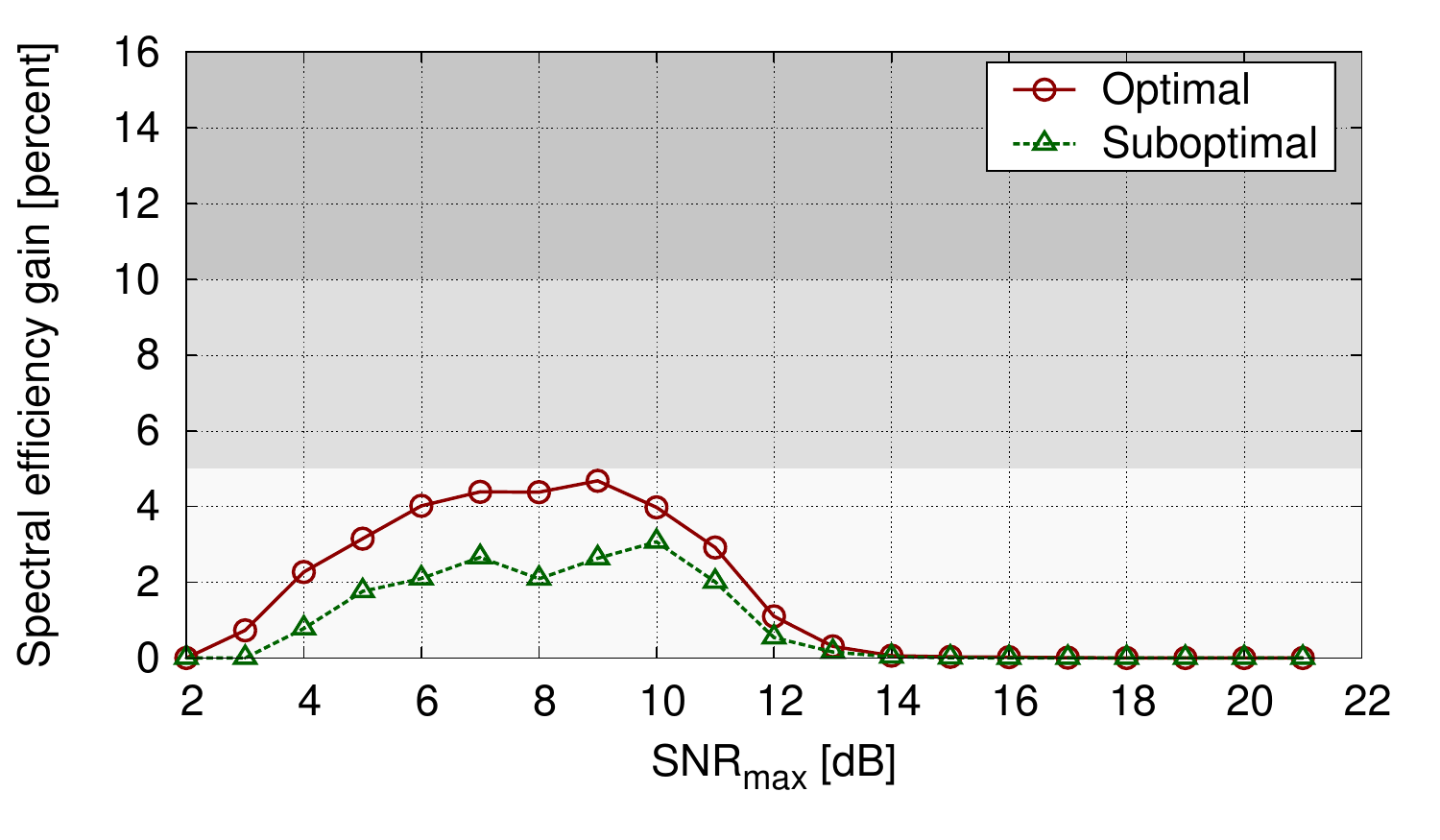}%
\label{psk}}}%
\centerline{\subfloat[Hierarchical 16-APSK (4 constellations)]{\includegraphics[width=0.45\textwidth]{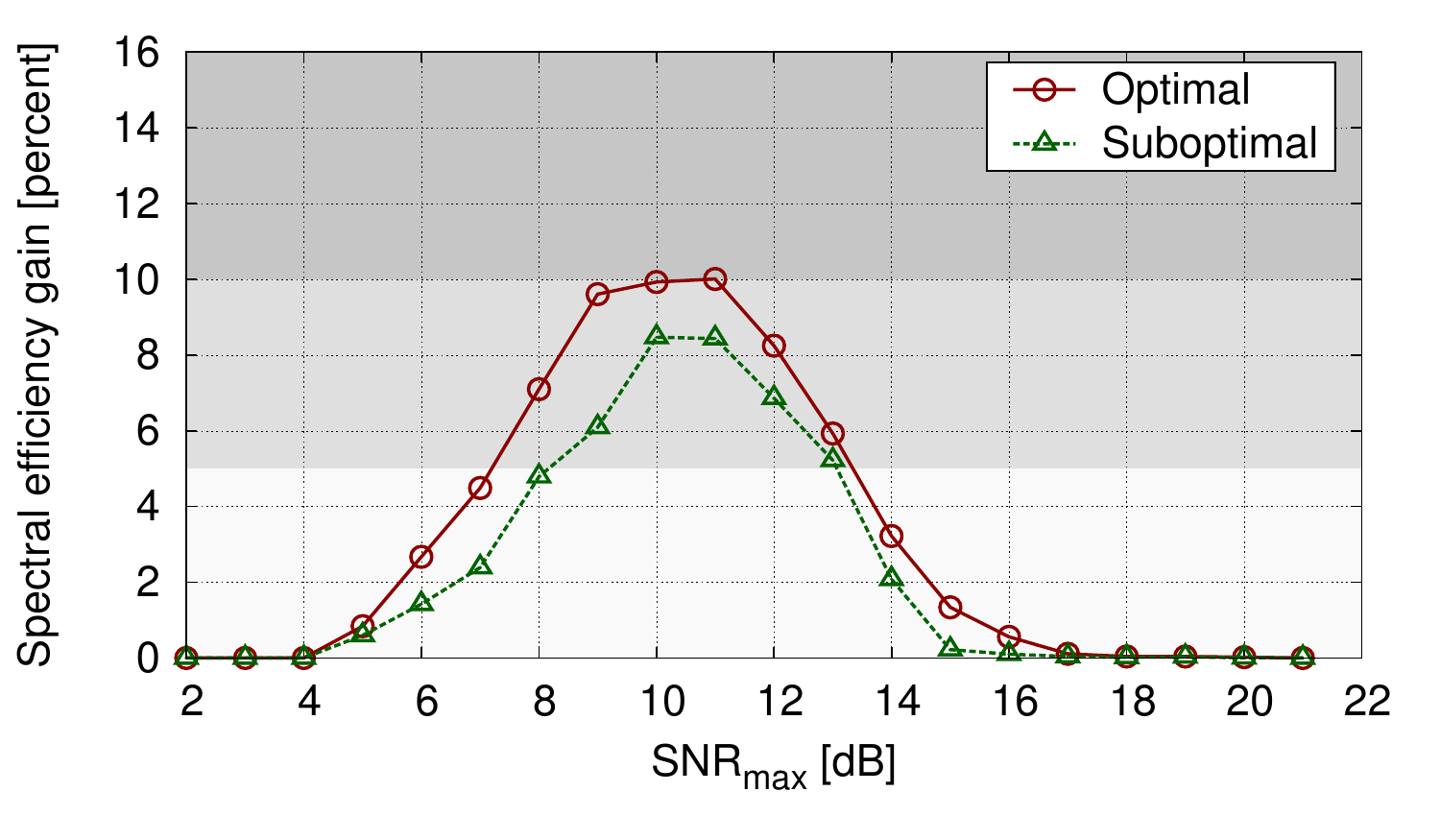}%
\label{16apsk}}%
\hfil
\subfloat[Hierarchical 32-APSK (5 constellations)]{\includegraphics[width=0.45\textwidth]{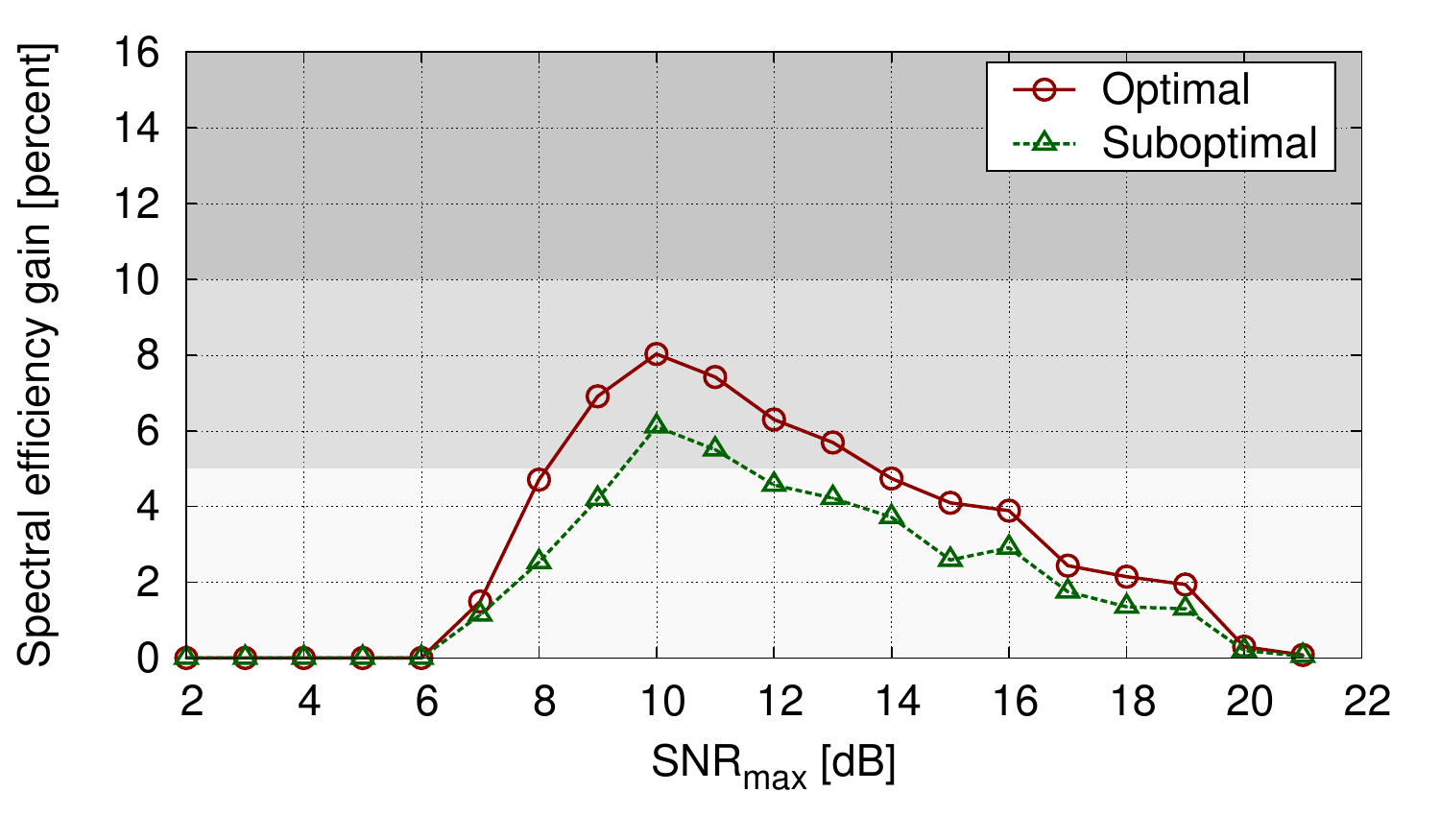}%
\label{32apsk}}}%
\caption{Performance of the optimal scheme based on linear programming for the different hierarchical modulations. We also plot the performance of the suboptimal solution proposed in \cite{icc14}. The spectral efficiency gain is measured relative to the reference scenario without hierarchical modulation.}
\label{perf}
\end{figure*}

\textbf{Results.}
\figurename~\ref{perf} presents the average spectral efficiency gains over 100 simulations. The results are shown separately for the different hierarchical modulations, enabling to visualise how each one affects performance. Moreover each subfigure has the same axes in order to ease the comparison and we also highlight three regions where the gains are less than 5\%, between 5\% and 10\%, or more than 10\%. 

Firstly, it is worth noting that the results for the suboptimal schemes are consistent with previous works, especially \cite[Fig.~9]{broad13} and \cite[Fig.~8]{icc14}.

Secondly, the results point out that the optimal scheme (based on linear programming) provides some improvements compared to the suboptimal one. For instance, considering the hierarchical QPSK with $\text{SNR}_{\text{max}}$ equals 0 dB, the performance increases from 12\% to 15\%. Even if we only notice a slight improvement in many cases, the optimal transmission rate is now known for our framework. Moreover, the linear program proposed in this paper may be adapted to other applications, such as bit division multiplexing as discussed in Section~\ref{part5}.

Then we remark that the performance strongly depends on the modulations. Each modulation is effective in a given SNR range. Indeed, the hierarchical QPSK enables a gain larger than 5\% for $\text{SNR}_{\text{max}}$ below 6 dB, while the 16 and 32-APSK performs well for better channel qualities (roughly, $\text{SNR}_{\text{max}}$ between 7 and 14 dB). Moreover, the gains are very different: the QPSK obtains a performance improvement up to 15\%, whereas the 8-PSK never exceeds 5\%. However, it may be possible to better optimise each constellation geometry; this is part of future work. 

With the current constellations, the hierarchical 8-PSK does not deserve to be kept as there is always another constellation, either the QPSK or the 16-APSK, that performs better. 

We originally designed the hierarchical 32-APSK to provide throughput improvement when the 16-APSK performance drops, i.e., for $\text{SNR}_{\text{max}}$ between 14 and 18 dB. The objective is not entirely reached as the observed gain is less that 5\% in the desired zone. For larger SNR, it is not possible to improve the transmission rate with the hierarchical 32-APSK as almost all the receivers decode the (non-hierarchical) 32-APSK with the highest code rate as previously mentioned. In that case the solution is to consider higher order constellations as the 64, 128 or 256-APSK. These modulations are now part of the DVB-S2X standard, an extension of DVB-S2.  

Lastly, we give a quick word about the link unavailability. This parameter is only interesting for low $\text{SNR}_{\text{max}}$ values as the beam coverage is total when all the receivers experience good channel conditions. In all our simulations, the unavailability stays below 3\%. If we consider $\text{SNR}_{\text{max}}$ values smaller than 2 dB, the unavailability becomes unsuitable for broadcast systems. For instance, the coverage is below 60\% when $\text{SNR}_{\text{max}}$ equals 0 dB as many receivers are not able to decode any transmitted stream due to the parameters in the standard.

\section{Related work and discussion}\label{part5}

\textbf{Related work.} Hierarchical modulation has many applications as already mentioned in the introduction. Although it has been introduced to increase the transmission rate of broadcast channels, only few works investigate to that end.

We recently pointed out the advantage of using 2-layers hierarchical modulation to improve the spectral efficiency in modern satellite system \cite{broad13, icc14}. We designed novel non-uniform constellations and studied their performance for a DVB-S2 system similar to the one presented in Section~\ref{part4}. 

In our previous works, the spectral efficiency optimisation was as follows: during the transmission, the source  communicates either with one receiver or with a pair of receivers to transmit with 2-layers hierarchical modulation. This generates a partition of the set of receivers. For each pair, we compute the best transmission rate offered to both receivers using hierarchical and non-hierarchical modulations. Finally, time sharing enables to equalise the spectral efficiency between all the receivers. This scheme also improves the performance but it is not optimal as shown in \figurename~\ref{perf}. The following example gives a geometric insight of why it is suboptimal.

\begin{figure}[!ht]
\centering
\includegraphics[width = 0.7\columnwidth]{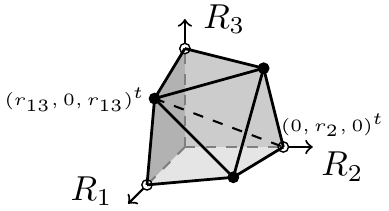}
\caption{Example of transmission rates with three receivers. The filled (resp. unfilled) circles correspond to hierarchical (resp. non-hierarchical) modulations achievable rate vectors.}
\label{achievable_rate}
\end{figure}

We consider a case with three receivers. \figurename~\ref{achievable_rate} illustrates the set of achievable rate vectors $(R_1,R_2,R_3)^t$ where $R_i$ is the transmission rate of receiver $i$ ($1 \leqslant i \leqslant 3$). There are three vectors obtained with non-hierarchical modulations and three with hierarchical modulations. The optimal rate corresponds geometrically to the intersection of the convex hull of all the achievable rate vectors with the line $R_1=R_2=R_3$ (to offer the same spectral efficiency to all the receivers). For the suboptimal scheme, we assume that receiver 1 is paired with receiver 3 and some transmission parameters (using hierarchical modulation) enable a rate vector of the form $(r_{13}, 0, r_{13})^t$. In other word, receiver 1 and 3 have a spectral efficiency of $r_{13}$. For its part, receiver 2 has a spectral efficiency equals to $r_2$ (using non-hierarchical modulation). Time sharing enables to equalise the transmission rates between each receiver and the final time-averaged spectral efficiency is
\begin{equation}
\underbrace{\frac{r_2}{r_2+r_{13}}}_{t_{13}} \begin{pmatrix} r_{13} \\ 0 \\ r_{13} \end{pmatrix} + 
\underbrace{\frac{r_{13}}{r_2+r_{13}}}_{t_2} \begin{pmatrix} 0 \\ r_2 \\ 0 \end{pmatrix} = 
\frac{r_2r_{13}}{r_2+r_{13}} \begin{pmatrix} 1 \\ 1 \\ 1 \end{pmatrix}.
\end{equation}
The terms $t_{13}$ and $t_2$ are some time sharing coefficients. The black dashed line in \figurename~\ref{achievable_rate} represents the achievable transmission rates of time sharing between $(r_{13}, 0, r_{13})^t$ and $(0, r_2,0)^t$. This line is inside the convex hull explaining why our previous scheme was suboptimal.

About the spectral efficiency optimisation, the source has many ways to partition the receivers and we suggested few heuristics with good performance \cite{broad13, icc14}. Later, we showed that the partition achieving the best transmission rate is the solution of an assignment problem, for which efficient algorithms exist such as the Hungarian method \cite{el14}.

Finally, even if the solution based on partitioning the receivers is suboptimal (in terms of spectral efficiency), it does not require intensive computation and can be done in real-time. This is of particular interest for modern satellite standards.

\textbf{Discussion.} First of all, the extension to $i$-layers hierarchical modulation ($i\geqslant3$) is immediate. Indeed, the fact that rate vectors can now have $i$ non-zero components does not modify the problem formulation in (\ref{canonical_form}). The main difference is from a practical point of view. Considering a system with $n$ receivers that implements $i$-layers hierarchical modulations, there are $\binom{n}{i}$ ways to pick $i$ receivers among $n$ to transmit with the corresponding hierarchical modulation. Thus the number of input rate vectors to the optimisation problem varies with the number of layers. This leads us to our second point concerning the time complexity to get the optimal solution.

The (time) complexity is an important factor if we want to use the optimal scheme in real-time. It depends on the algorithm used by the solver and the size $k$ of the rate vectors set. As the number of receivers and the transmission parameters affect $k$, the complexity is ultimately dependent on these system parameters. However the two critical values are the number of receivers and the maximum number of layers for hierarchical modulation. In our current framework, we consider a broadcast area of 500 receivers, the system implements 2-layers hierarchical modulation and the solver relies on an interior-point algorithm. Each simulation requires few minutes and thus can not operate in real-time. To tackle this limitation, we will investigate as future work some heuristics to speed up the complexity. Another solution is to rely on a suboptimal scheme as proposed in \cite{broad13, icc14}. 

In this paper, we focus on offering the same time-averaged spectral efficiency to all the receivers. This practical example enables to evaluate the potential of combining time sharing with hierarchical modulation. Nevertheless some scenarios require to include rate constraints. For instance, we may consider a system with premium and non-premium receivers where premium receivers pay to obtain a better throughput. The introduction of rate constraints only requires to modify the last column of the matrix defined in (\ref{A}). 

Recently, a novel channel resource allocation has been introduced \cite{6509443}. This scheme, called bit division multiplexing, extends the multiplexing from symbol level to bit level. In their work, the authors optimise the transmission rate or the decoding threshold when communicating two services but the case with $n\geqslant3$ services is not treated. The framework proposed here can be adapted to optimise the transmission rate of bit division multiplexing by modifying the matrix and vectors defined in (\ref{canonical_form}).

Lastly, we focus on optimising the transmission rate with the same link unavailability as the reference system without hierarchical modulation. Another research direction is to study the trade-off between throughput increase and coverage extension. Indeed, the proposed modulations enable to have lower decoding thresholds than the ones in the DVB-S2 standard.

\section{Conclusion and future work}\label{part6}

We study a DVB-S2 broadcasting system that combines VCM with hierarchical modulation. We show that the optimal transmission rate is the solution of a linear programming problem. The optimal scheme achieves significant gains, around 10\%, for a large range of channel quality.

Future work will explore the issues raised in the discussion part. Moreover, our framework enables to obtain the optimal spectral efficiency for a given set of modulations. Even if we designed the constellations considered in this paper, a step further is to include the design in the optimisation to obtain the optimal constellation parameters.

\section*{Acknowledgment}
The author wishes to thank J\'er\'emy Barbay and the ACGO (algorithms, combinatorics, game theory and optimization) group for useful discussions.

\ifCLASSOPTIONcaptionsoff
  \newpage
\fi

\nocite{*}
\bibliographystyle{IEEEtran}
\bibliography{biblio}

\end{document}